\documentstyle[epsfig,11pt]{article}
\begin{document}
\setlength{\baselineskip}{7mm}
\begin{center}
{\large\bf
On the   field theoretical formulation
of the electron-proton scattering in the
Coulomb and Lorentz gauges. }
\end{center}
\vspace{1cm}
\noindent{\em A.I.\ Machavariani}

\vspace{3pt}
\noindent{ Joint\ Institute\ for\ Nuclear\ Research,\ LIT\ Dubna,\ Moscow\
region\ 141980,\ Russia\\
and High Energy Physics Institute of Tbilisi State University,
University str.  9 }\\
{\em Tbilisi 380086, Georgia }

\vspace{0.5cm}

\begin{abstract}

The relativistic three dimensional (3D)
Lippmann-Schwinger-type  equations 
for the $ep$ scattering amplitude is derived based on 
 unitarity condition in the usual quantum electrodynamic (QED).
The $ep$ scattering potential $V_{e'N',eN}$
consists of the leading one off mass shell photon exchange part 
 and the nonlocal multi-particle exchange potential . 
Unlike to the other field-theoretical equations, both protons 
in the unitarity condition and in $V_{e'N',eN}$ are on mass shell. 
Therefore in this approach  are not required the 
multi-variable  input photon-nucleon vertexes
with the off mass shell nucleons.

 In the present formulation the standard  
 leading   one photon exchange potential $V_{OPE}$ 
is generated by the  canonical equal-time anti commutator
between the electron source and the interacted electron  fields 
which are sandwiched by the one nucleon asymptotic states.
This anticommutator  is calculated in the Coulomb and Lorentz 
gauges, where only the  transverse parts of the photon fields
are quantized.  It is shown, that the leading one photon exchange potential
 $V_{OPE}$ in the Coulomb and Lorentz gauges coincide.
The complete set of the next to leading order terms which are generated
by the static electric (Coulomb) interaction 
 are exactly reproduced.



\end{abstract}

\newpage

\begin{center}
{\bf{1. Introduction}}
\end{center}

\vspace{0.25cm}

Presently the elastic $ep$ scattering 
in  energy  region up to few $GeV$ is well described
by the leading order one photon exchange (OPE) model.
In this energy  region the various observables for the $ep$
scattering reactions 
were studied within the next of leading order two photon exchange 
(TPE) models in order to check the accuracy of the OPE model.
The OPE model was satisfactorily justified by description 
of the ratio of the electric and magnetic form factors
of the proton, angular dependence of the polarisation observables,    
electro-production of the pions and resonances, the ratio of the
cross sections of the 
$e^-p$ and $e^+p$ reactions, electromagnetic form factors of the neutron
and other observables in the high precision experiments 
\cite{Melnichouk1,Vanderhaeghen1,Vanderhaeghen2}. 
The important uncertainty in the calculations of the $ep$ scattering 
within the TPE models arise from off shell extension of
the $\gamma^*pp$ vertexes \cite{AMT} because the loop diagrams contain the
$\gamma^*pp$ vertexes  with off mass shell nucleons which
depends on the two or three variables. Therefore there was used
the additional models in order to construct or simulate the multi-variable
$\gamma^*pp$ form factors.

An other reason for investigation of the high order 
power series expansion in $e^2$ of the $ep$ scattering amplitude
is that  60 years ago Dyson argued 
(see  \cite{Dyson}, ch. 9.4 in \cite{IZ}  and ch. 20.7 in\cite{Wei2})
 that the perturbation series 
in QED are divergent after renormalization of the mass 
and charge.
Therefore, there is necessary to use the special methods for
calculations of the two-photon and multi-photon exchange effects.  
For instance, the perturbative series in QED 
 can be calculated via the special asymptotic expansions \cite{Suslov} 
or using 
the mathematical methods without power series expansion in $e^2$.
One of the non perturbative 
approaches based on the relativistic field theoretical
generalization of the Lippmann-Schwinger  
equations $A=V+VG_oA$, where one can determine $(G_o-V)^{-1}$ and
the full $ep$ scattering amplitude $A=V(1-G_oV)^{-1}$ even if the perturbative 
series $A=V+VG_oV+VG_oVG_oV+...$ are divergent.

The 3D quasipotential reductions 
of the 4D Bethe-Salpeter equation for the Hydrogen-type systems 
are used in \cite{Eades,Karschenboim, Bodwin}.
In these approaches the 3D relativistic Lippmann-Schwinger type
equations were applied to the high order 
perturbation terms for the energy levels. But the calculation are
performed based on the eigenfunctions of the Schr\"odinger 
equation with the Coulomb potential  
or the eigenfunctions of the Dirac equation in the external 
Coulomb field.   
The solution of the relativistic Dirac equation was applied also
 for calculation of the ion-atom scattering \cite{Eichler}.


This paper deals with the new kind  
3D Lippmann-Schwinger-type field-theoretical equations for
the $ep$ scattering amplitude $A_{e'N',eN}$   within the usual 
quantum electrodynamic (QED), where
the protons in the corresponding potential $V_{e'N',eN}$
are on mass shell.
These 3D Lippmann-Schwinger equation allow to build 
$ep$ discrete spectrum and the $ep$  elastic scattering amplitudes based on 
the same potentials.
The present approach based on the 3D field-theoretical
unitarity condition and  is free from the 3D ambiguities which appears 
by the 3D (quasipotential) reductions of the 4D Bethe-Salpeter equations.  
Unlike to the 4D Bethe-Salpeter 
equations and their 3D quasipotential reductions, the 
 potential  of the suggested equations
is constructed from  the one-variable vertexes 
 with on mass shell protons.
The off mass shell intermediate particle exchange part 
in the suggested approach 
is generated by the canonical equal time anti commutators which is 
sandwiched by the asymptotic one proton states. 
This part of the potential generates the leading 
OPE potential.
These  equal-time canonical anti commutators we shall derive  
in the Coulomb and Lorentz gauges.

The considered method was firstly applied to the low energy
$\pi N$ scattering by Chew and Low \cite{ChewLow} and was developed 
by Banerjee and coworkers \cite{Ban}. The analytical 
model-independent linearization of these relativistic
quadratically nonlinear equations was done in \cite{MR,MCH,M}, where they 
were  applied to  descriptions
of the low energy  $\pi N $ and $N N$ phase shifts.
The leading order one-particle exchange diagrams in this approach  for 
the multichannel reactions $ep-e'p'\gamma'$ and $\gamma p-\gamma.\pi' p$
in the $\Delta$-resonance region was studied in \cite{MBF,MF,MFPR,MFJPG}.

The present paper consists of the five sections. In the next section
the general unitarity condition for the $ep$ scattering amplitudes
with the on mass shell protons and the corresponding 3D Lippmann-Schwinger 
type equation are considered. 
The calculation of the OPE potential through the equal time
canonical anti commutators in the Coulomb gauge is performed in section 3.  
In the section 4
the same anti commutators and the potentials are calculated in the Lorentz 
gauge and  the exact relationship of these expressions in
the Lorentz and in the Coulomb gauges are suggested. 
The Summary is  given in section 5.

\vspace{0.25cm}

\begin{center}
{\bf{2. Unitarity condition and corresponding 3D relativistic equation  
within the time ordered relativistic field-theoretical approach}}
\end{center}

\vspace{0.25cm}

In the  quantum field theory \cite{IZ,BD}
the  elastic $ep$ scattering amplitude is

$${\cal A}_{e'N',eN}
=-< out;{\bf p'}_N|\eta_{\bf p_e'}(0)|{\bf p}_e{\bf p}_N;in>
\eqno(2.1a)$$
where $(p_e^o\equiv E_{\bf p_e}=\sqrt{{\bf p}^2_e+m_e^2},{\bf p_e}$),  
$(p^o_N\equiv E_{\bf p_N}=\sqrt{{\bf p}^2_N+m_N^2},{\bf p_N})$ and 
$({p'}_e^o\equiv E_{\bf p_e'}=\sqrt{{\bf p'}_e+m_e^2}, {\bf p_e'})$,  
$({p^o}'_N\equiv E_{\bf p'_N}=\sqrt{{\bf p'}_N+m_N^2}, {\bf p_N'})$
denote the energy-momentum  of the on mass shell electron and proton 
in the initial and final 
states. $\biggl(i\gamma_{\mu}{{\partial}/{\partial x_{\mu}}}-m_e
\biggr)\psi_{e}(x)=\eta_{e}(x)$ stand for the source operators 
of the electron,  and
$$\eta_{\bf p_e'}(x)={\overline u}({\bf p_e}')\eta(x)\eqno(2.1b)$$
The normalizations and designations of the symbols 
in above expressions
are taken from \cite{IZ}. 
For the sake of simplicity  the spin indexes in the 
expressions (2.1a,b) and in the Dirac spinor  $u({\bf p_{e}})$
are omitted.

In the amplitude (2.1a) the final electron is extracted from the asymptotic
"out" states and it depends on ${\bf p}'_e$ only  through 
${\overline u}({\bf p_e}')$. Therefore, the 4-momenta of the final 
electron $e'$ can be chosen in such a way that $e'$ 
is on energy shell but off mass shell, i.e $(p'_e)_{\mu}=
(p_e)_{\mu}+(p_N)_{\mu}-(p'_N)_{\mu}$ and $(p'_e)^2\ne m^2_e$.
According  to the reduction formulas \cite{BD} the $ep$ scattering amplitude 
 can be represented as 

$${\cal A}_{e'N',eN}=< out;{\bf p'}_N|
\biggl(\Bigl\{\eta_{\bf p_e'}(0),{{b}}^+_{\bf {p_e}}(0)\Bigr\}
-i\int d^4x e^{-ip_ex}\theta(-x_o)
\Bigl\{\eta_{\bf p_e'}(0),{\overline \eta}_{\bf p_e}(x))\Bigr\}\biggr)
|{\bf p}_N;in>,\eqno(2.2) $$
where the curly braces 
denote the anticommutator of the corresponding operators, 
$b^+_{{\bf p}_e}(x_o)$ transforms 
into corresponding creation or annihilation operator in the  
asymptotic $out(in)$ region $lim_{x_o\to\pm \infty}^{weakly}b^+_{{\bf p}_e}(x_o)
\longrightarrow b^+_{{\bf p}_e}\Bigl(out(in)\Bigr)$
 and 

$$b^+_{{\bf p}_e}(x_o)=\int d^3x 
e^{-iE_{\bf p_e}x_o+i{\bf p_{e}x}}{\overline \psi}_{e}(x)\gamma_o u({\bf p_e});
\ \ \ \ \ \ \  
{\stackrel{o}{b}}^+_{\bf {p_e}}(x_o)\equiv
 { {\partial b^+_{{\bf p}_e}(x_o) }\over {\partial x_o}}=
i\int d^3x e^{-iE_{\bf p_e}x_o+i{\bf p_{e}x}}{\overline \eta}_{\bf p_e}(x).
\eqno(2.3)$$
For derivation of (2.2)
the identity
$b^+_{{\bf p}_{e}}(in)=b^+_{{\bf p}_{e}}(0)-\int dx_o\theta(-x_o)
{\stackrel{o}{b}}^+_{\bf {p_e}}(x_o)$ was used.

Substitution the completeness condition of the asymptotic in-states
$$\sum_n |n;in><in;n|={\sf 1}\eqno(2.4)$$ 
between the source operators in (2.2)
and integration over $x$ yields 

$$-< out;{\bf p'}_N|\eta_{\bf p_e'}(0)|{\bf p}_e{\bf p}_N;in>=
< out;{\bf p'}_N|\biggl\{\eta_{\bf p_e'}(0),{{b}}^+_{\bf {p_e}}(0)
\biggr\}|{\bf p}_N;in>$$
$$+\sum_{n=H,ep,ep\gamma,...}
<out;{\bf p'}_N|\eta_{\bf p_e'}(0)|n;in>
{{(2\pi)^3\delta({\bf P_n-p_e-p_N})}\over {P_o-P_n^o+i\epsilon}}
<in;n|{\overline \eta}_{\bf p_e}(0)|{\bf p}_N;in>$$
$$+\sum_{m={\overline e}p,{\overline e}\gamma p...}
<out;{\bf p'}_N|{\overline \eta}_{\bf p_e}(0)|m;in>
{{(2\pi)^3\delta({\bf P_m+p_e-p_N'})}\over {
E_{\bf p_e}-E_{\bf p_N'}+P_m^o}}
<in;m| \eta_{\bf p_e'}(0)|{\bf p}_N;in>,
\eqno(2.5)$$
where $P_o=E_{\bf p_e}+E_{\bf p_N}$ and   $P_o'=E_{\bf p_e'}+E_{\bf p_N'}$
are the  energies  of the $ep$ system in the initial and final states, 
$P_n^o$ and ${\bf P}_n$ stand for the energy and momentum of the 
intermediate $n$-particle states, 
$H$ and ${\overline e}$ denote the on mass shell intermediate
 $ep$ bound state (Hydrogen)
and positron.

If all of the fourth components of the photon field $A_{\mu}(x)$
are independent and quantized, then we have the following canonical commutation 
relations for the interacted electron
and photon fields \cite{IZ}

$$\Bigl[{{\partial A_{\mu}(x_o,{\bf x})}\over {\partial x_o}},
A_{\nu}(x_o,{\bf y})\Bigr]=
ig_{\mu\nu}\delta({\bf x-y})\eqno(2.6a)$$
$$\Bigl[A_{\mu}(x_o,{\bf x}),A_{\nu}(x_o,{\bf y})\Bigr]=0;\ \ \ 
\Bigl[\psi(x_o,{\bf x}),A_{\nu}(x_o,{\bf y})\Bigr]=0;
\eqno(2.6b)$$
$$\Bigl\{{\psi}(x_o,{\bf x}),\psi^+(x_o,{\bf y})\Bigr\}=
\delta({\bf x-y})\eqno(2.6c)$$ 
that allow to determine the first term in (2.2)
$${ Y}_{e'N',eN}=
< out;{\bf p'}_N|\biggl\{\eta_{\bf p_e'}(0),{{b}}^+_{\bf {p_e}}(0)
\biggr\}|{\bf p}_N;in>\eqno(2.7a)$$
as the usual leading one photon exchange potential
$${ Y}_{e'N',eN}=V_{OPE}
=e{\overline u}({\bf p_e'})\gamma_{\mu}u({\bf p_e})
{1\over t}_N<out;{\bf p}_N'|J^{\mu}(0)|{\bf p}_N;in>,\eqno(2.7b)$$
where $t_N=({p_N^o}'-p_N^o)^2-({\bf p'_N-p_N})^2$   and
$J^{\mu}(x)$ denotes the  photon source  
$$J^{\mu}(x)=e{\overline \psi}_e(x)\gamma^{\mu}\psi_e(x)
+e{\overline \psi}_N(x)\gamma^{\mu}\psi_N(x)+...\eqno(2.8)$$
 where $...$ stands for the 
sources of the intermediate particles.

The $\gamma^*p'p$-vertex $<out;{\bf p}_N'|J^{\mu}(0)|{\bf p}_N;in>$
in (2.7b) with the off mass shell proton 
is determined via the usual Dirac
electric and magnetic form factors of the proton     
$$<out;{\bf p}_N'|J^{\mu}(0)|{\bf p}_N;in>=
{\overline u}({\bf p_N'})\Bigl(\gamma^{\mu}F_1(t_N)+
i{{ \sigma^{\mu\nu}(p'-p)_{\nu}}\over {2m_N}}F_2(t_N)\Bigl)
\eqno(2.9)$$

 The graphical representation of the   one off mass shell photon exchange potential
$V_{OPE}$ (2.7b) is given in Fig. 1. In this diagram the shaded circle corresponds to
the $\gamma^*NN$ vertex  (2.9) and the dot between the 
electron-photon lines relates to the 
$\gamma^*ee$ vertex in the tree approximation.

\vspace{0.15cm}

\begin{figure}[htb]
\centerline{\epsfysize=145mm\epsfbox{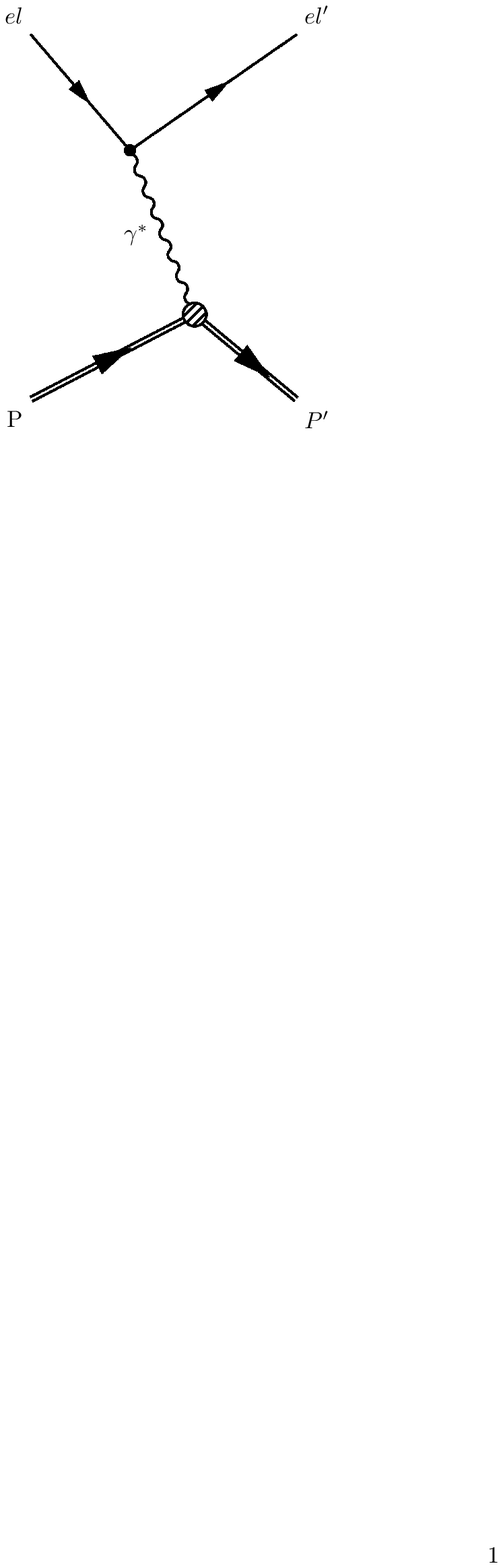}}
\vspace{-10.5cm} 
\caption{{\protect {\em  The $t$-channel
 one photon exchange term (2.7b) with off mass shell proton.
}}}
\end{figure}

\vspace{0.15cm}

The canonical commutation relations (2.6a,b) requires the
the indefinite metric according to 
Gupta-Bleuer approach\cite{IZ}.
In order to avoid this complication 
in the next sections we shall consider the 
Lorentz and Coulomb gauges where only the transversely parts of the
photon fields are independent and quantized.

The vertexes  of the transition in the intermediate 
$n$ and $m$ particle states in (2.5) consists of the connected and
disconnected parts. The cluster decomposition procedure \cite{alf,Ban}
allow to represent the particle exchange paths in (2.5) through the
connected vertexes. Then the second and third terms in (2.5) generate
the eight expression (A.1a)-(A.1h) in Appendix A with $n=H,ep,ep\gamma,...$.
The on mass shell particle exchange terms   (A.1a)-(A.1h) with the
simplest  two and three body intermediate states
are depicted  in Fig. 2.
These terms have different chronological sequences of the absorption and
 emission of the external electrons and nucleons. 
In particular, the $s$-channel diagram in Fig. 1A 
the initial $eN$ state transfers into intermediate
 $ \gamma e'' N''$ state which afterwards transforms into final 
$e'N'$ state. 
In Fig. 1B  the photon-electron
intermediate state $ \gamma e''$ is generated by
the $eN\to \gamma e''N$ transition amplitude and 
 after truncation of the photon 
$ \gamma e''$ transforms  into  $e'$. In Fig.1C firstly electron radiate
photon and afterwards  the intermediate $ \gamma e''$ state
is truncated via  the amplitude   
$\gamma e''N\to e'N'$.  The diagram in Fig. 1D describes
the chain of reactions, where firstly the initial electron 
generates the $e''{\overline N} N'$ state with the final nucleon $N'$
and afterwards this intermediate state together with the initial
nucleon $N$ produces the final electron $e'$.
The $u$-channel terms in the
diagrams in Fig. 1E,F,G.H are obtained from the $s$-channel terms in
 Fig. 1A,B,C,D via crossing of the external electrons.
Therefore, in Fig. 1E,F,G.H firstly is radiated the final electron and 
afterwards is truncated the initial electron. Consequently the chine of
the reactions in Fig. 1E,F,G.H contains the intermediate anti-electron
(positron) state ${\overline e}$. 
Thus  any $s$-channel diagram  have the corresponding 
antiparticle exchange so called $Z$ diagrams \cite{alf}.

\vspace{0.15cm}


\begin{figure}[htb]
\centerline{\epsfysize=180mm\epsfbox{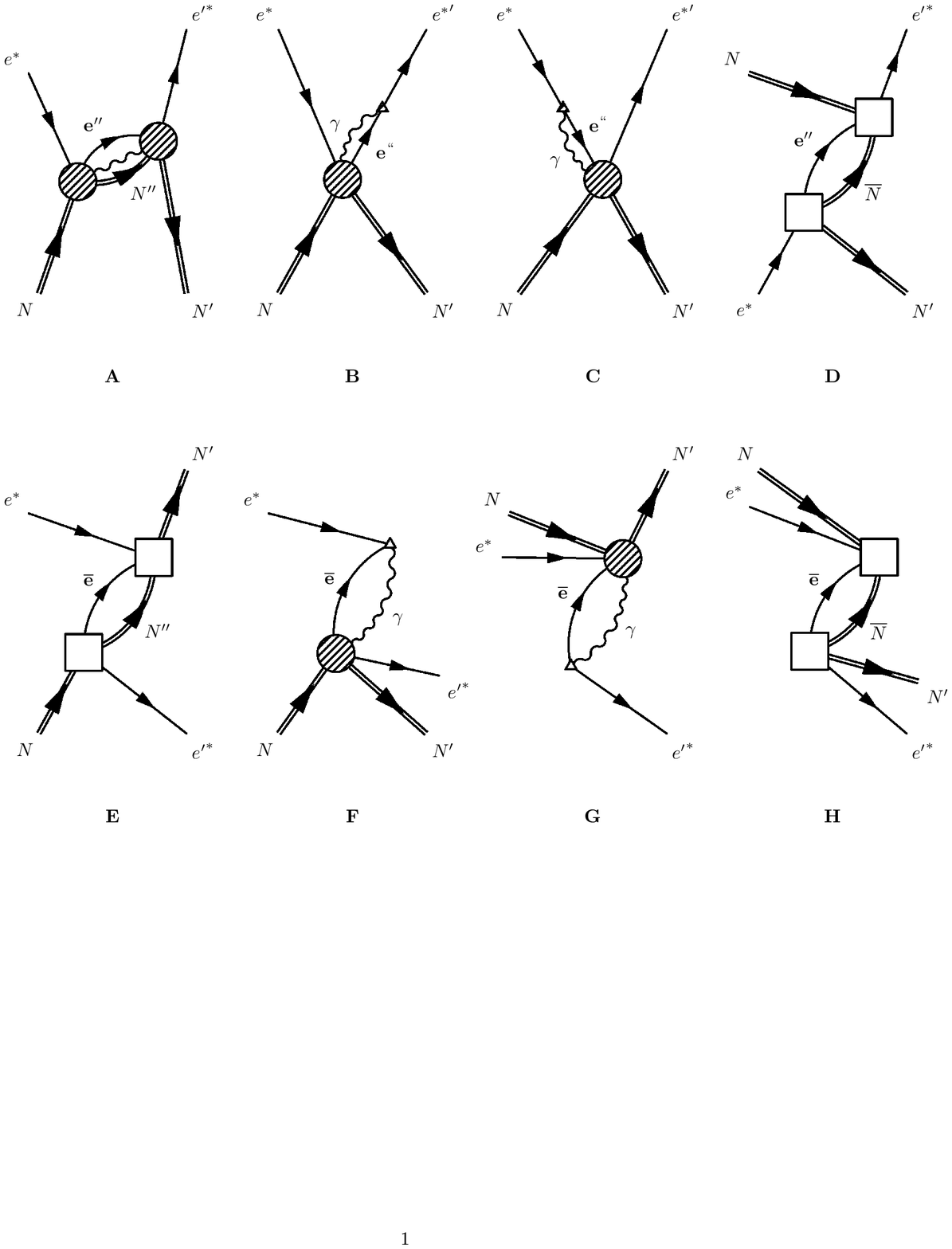}} 
\vspace{-5.5cm} 
\caption{{\protect {\em On mass shell particle exchange 
time ordered diagrams with the off mass shell electrons.
The dashed circle denotes the five particle
$ep\leftrightarrow ep\gamma$ 
transition amplitudes and their crossings. 
Square stands for the four point 
amplitudes $ep\leftrightarrow ep$  and their crossings. 
  (D) and (H) presents the diagrams with the intermediate anti nucleon.
(E), (F), (G) and (H) are the $u$-channel diagrams   with the intermediate 
anti electron exchange.
}}}
\end{figure}

\vspace{0.15cm}

The particle in the intermediate states in the expressions (A.1a)-(A.1h)
are on mass shell. Therefore unlike to the 4D formulations
in (A.1a)-(A.1h) do not contain the  energy and charge renormalization
diagrams and proton vertex correction diagrams.

  It is  convenient to consider separately 
 the $s$-channel  exchange terms with the intermediate states $n=H,ep$

$${\cal A}_{e'N',eN}={ Y}_{e'N',eN}+{\cal W}_{e'N',eN}+
\sum_{H} {\cal A}_{e'N',H}{1\over {P_o-P_H^o}}{{\cal A}^+}_{H,eN}
+\sum_{e"N"} {\cal A}_{e'N',e"N"}{1\over {P_o-P_{e"N"}^o}+i\epsilon}
{{\cal A}^+}_{e"N",eN},
\eqno(2.10)$$
where ${\cal W}_{e'N',eN}+$ is determined in (A.1a)-(A.1h) without
on shell $n=H,ep$ exchange terms. 

It is easy to check, that the $ep$ scattering amplitude ${\cal
  A}_{e'N',eN}$ in (2.5) and in (2.10)
satisfies the general unitarity condition with all possible $n$-particle
$s$-channel intermediate states
$${\cal A}_{e'N',eN}-{\cal A}_{e'N',eN}^+=\sum_{n=ep,ep\gamma,'...}
{\cal A}_{e'N',n}\biggl[
{{(2\pi)^3\delta({\bf P_n-p_e-p_N})}\over {P_o-P_n^o+i\epsilon}}-
{{(2\pi)^3\delta({\bf P_n-p'_e-p'_N})}\over {P'_o-P_n^o-i\epsilon}}
 \biggr]{\cal A}_{en,n}^*
\eqno(2.11)$$
 
The analytic linearization of the 3D time-ordered
and quadratically nonlinear equation (2.10) 
 can be performed in the same way   
as for the elastic $\pi N$ or $NN$ scattering in \cite{MR,MCH,M}
or for the reaction $\pi N-\gamma' N'$ \cite{MBF}-\cite{MFJPG}.
The resulting relativistic Lippmann-Schwinger type is

$${\cal T}({\bf p',p},P_o)={\cal U}({\bf p',p},P_o)+\int 
{\cal U}({\bf p',q},P_o){{ d^3{\bf q}}\over {P_o-
E_{\bf p_e'}-E_{\bf p_N'}+i\epsilon} }{\cal T}({\bf q,p},P_o)\eqno(2.13)$$
 
where the on shell ${\cal T}({\bf p',p},P_o)$  coincides with 
the on shell amplitude ${\cal A}_{e'N',eN}$ (2.1a)

$$\Biggl[{\cal T}({\bf p',p},P_o)
=-< out;{\bf p'}_N|\eta_{\bf p_e'}(0)|{\bf p}_e{\bf p}_N;in>\biggr]^{on\ \ \ shell},
\eqno(2.14)$$
if ${\cal U}({\bf p',p},P_o)$ is constructed through the 
inhomogeneous term of (2.10) as

$${\cal U}({\bf p',q},E)=Y_{e'N',eN}({\bf p',p})+
{\sf A}_{e'N',eN}({\bf p',p})+
E{\sf B}_{e'N',eN}({\bf p',p})\eqno(2.15a)$$ 

where

$${\cal W}_{e'N',eN}({\bf p',p})=
{\sf A}_{e'N',eN}({\bf p',p})+
P_o'{\sf B}_{e'N',eN}({\bf p',p}),\eqno(2.15b)$$

The exact form of ${\cal W}_{e'N',eN}$, ${\sf A}_{e'N',eN}$ and 
${\sf B}_{e'N',eN}({\bf p',p})$ is  given in Appendix A,  

In the half off energy region ${\cal A}_{e'N',eN}$ (2.1a) is determined through 
 ${\cal T}({\bf p',p},P_o')$ as 
$$-< out;{\bf p'}_N|\eta_{\bf p_e'}(0)|{\bf p}_e{\bf p}_N;in>=
{\cal U}({\bf p',q},P'_o)+\int 
{\cal U}({\bf p',q},P_o'){{ d^3{\bf q}}\over {P_o-
E_{\bf p_e'}-E_{\bf p_N'}+i\epsilon} }{\cal T}({\bf q,p},P_o)\
\eqno(2.16)$$
where $P_o'=E_{\bf p_e'}+E_{\bf p_N'}$.


Thus the solution of the 3D  relativistic Lippmann-Schwinger-type equation 
(2.13) determine  the $eN$ scattering amplitude 
$-< out;{\bf p'}_N|\eta_{\bf p_e'}(0)|{\bf p}_e{\bf p}_N;in>$ (2.1a)
 according to (2.14) and  (2.16). 

The ${\cal S}$-matrix reduction formulas \cite{IZ,BD} allow to
 reproduce  the amplitude (2.1a) 
from the full Green function $<0|{\sf T}\Bigl(
\psi_e(0)\psi_N(x){\overline \psi}_e(y_e){\overline \psi}_e(y_N)\Bigr)|0>$
and the corresponding 4D Bethe-Salpeter amplitude.
Unlike to the 4D Bethe-Salpeter 
equations and their 3D reductions
 in the field-theoretical 3D equation (2.13) 
as input appear  the leading OPE potential (2.7b) with the  
usual one-variable form factors (2.9). Besides the solution 
of the 3D equation (2.13) satisfy the unitarity condition even 
after truncation of the intermediate on mass shell particle states in (A.1a)-(A.1h).
In the unitarity conditions (2.10) and the corresponding 3D Lippmann-Schwinger type 
equations nucleons are on mass shell. Therefore the quark-gluon degrees of freedom
can contribute only through the input OPE potential (2.7b) and the form factors (2.9).



\begin{center}
{\bf{3. 
Canonical commutation relations and  the Born term (2.7a) 
 in the Coulomb gauge. }}
\end{center}

\vspace{0.25cm}

In  the Coulomb gauge 
$${{\partial A^{C}_i(x)}\over{\partial x_{i}}}=0;\ \
 \ \ \ \ \ \ \ \ \ \ i=1,2,3\eqno(3.1)$$
as independent  are considered  the electron fields $\psi_e(x)$, $\psi_e^+(x)$ and
the transverse components of the photon field 
$$A^{C}_i\equiv A^{tr}_i=A_i-{1\over \Delta}{{\partial A_i}\over{\partial x_{i}}};
 \ \ \ \ \ \ \ \ \ \Delta={{\partial ^2}\over{\partial x_{i}\partial x^{i}}}  
\eqno(3.2)$$
The  conserved current $J^{\nu}(x)$ (2.8)
and the gauge condition (3.1) 
yields  the  equations of motion \cite{IZ,BD,Wei1}

$$\Box_x {A^C}_{i}(x) 
=J^{tr}_i(x)=J_i(x)-
{{\partial}\over {\partial x^{i}}} 
 {{\partial J^{k}(x)}\over {\partial x^{k}}} 
\eqno(3.3)$$

$$\Bigl(i\gamma^{\mu}{{\partial}\over{\partial x^{\mu}}}
-m_e\Bigr)\psi_e(x)= e\gamma ^{\mu} {A^C}_{\mu}(x)\psi_e(x)\eqno(3.4)$$

where the zero component of the photon field   ${A^C}_o$  is defined via 
the photon source (2.8) according to the Poison equation   

$$-\Delta  {A^C}_{o}(x)\equiv -{{\partial}\over {\partial x^{i}}} 
 {{\partial {A^C}_{o}(x)}\over {\partial x_{i}}} 
=J^{o}_i(x);\ \ \ \ \ 
 A^{C}_o(x)=\int {{ d{\bf x'}J_o(x_o,{\bf x'})}\over
{4\pi |{\bf x-x'|}} }
\eqno(3.5)$$
The equation of motion (3.3) does not contain the longitudinal part 
of the $A^C_i$ in virtue of  (3.1). 
Nevertheless in the electromagnetic Lagrangian 
$${\cal L}_{em}^{(C)}=-{1\over 4}F_{\mu\nu}F^{\mu\nu}+J_{i}^{tr}(A^{C})^i\eqno(3.6)$$
with   $F_{\mu\nu}(x)={{\partial A^{\nu}(x)}/{\partial x^{\mu}}}-
{{\partial A^{\mu}(x)}/{\partial x^{\nu}}}$ 
the longitudinal part of the electric field   
 $ {\bf E}^{l}_i=-{{ \partial A^C_o}/{\partial x_i}}$ appears
through the complete  electric fields
$${\bf E}_i=F_{io}
={\bf E}^{tr}_i+{\bf E}^{l}_i=-{ {\partial A_o}\over {\partial x^i}}-
{ {\partial A_i}\over {\partial x_o}};
\ \ \ \ \ \ \ \ \ \ {(\bf E^{tr})}_i=-{ {\partial A^C_i}\over {\partial x_o}}
\eqno(3.7)$$
Therefore one can separate the electrostatic (Coulomb) interaction term $J^oA^C_o$
in (3.6)  \cite{BD,Wei1}
$${\cal L}_{em}^{(C)}=-{1\over 2}\Bigl({{\bf E}^{tr}}\Bigr)^2-{1\over 2}{\bf B}^2
-J^{tr}_i{A^{C}}_i+J^oA^C_o\eqno(3.8a)$$
where ${\bf B}=-rot {{\bf A}^{C}}$ and $J^oA^C_o$ 
was generated by  ${\bf E}^{l}_i$
according to the relation
$$\int d^3{\bf x}({\bf E^l})^2=\int d^3{\bf x}A^C_oJ^o\eqno(3.8b)$$

The canonical commutation relations for $A^C_i$ \cite{BD,Wei1} are

$$\biggl[{ {\partial A^C_i(x_o,{\bf x})}\over{\partial x_o} },
A^C_j(x_o,{\bf y})\biggr]=
\delta_{ij}\delta({\bf x-y})-
{1\over {\Delta}}{{\partial^2}\over {\partial x_i\partial y_j}}
{1\over{|{\bf x-y}|}} \eqno(3.9)$$

The source  $\eta^C=e\gamma^o A^C_o \psi_e-e\gamma^iA_i^C\psi)1$ 
 in (3.4) and the canonical commutators (2.6c)
allow to represent the Born term (2.7a)  as
 
$${ Y}_{e'N',eN}^C\equiv Y_{I}^C-Y_{II}^C=\eqno(3.10a)$$
where
$$Y_{I}^C=e
{\overline u}({\bf p_e'})\gamma^{\mu}
< out;{\bf p'}_N|A^C_{\mu}(0)
\biggl\{\psi_e(0),{b}^+_{\bf {p_e}}(0)\biggr\}|{\bf p}_N;in>
=e{\overline u}({\bf p_e'})\gamma^{\mu}u({\bf p_e})
<out;{\bf p}_N'|A^C_{\mu}(0)|{\bf p}_N;in>
\eqno(3.10b)$$
$$Y_{II}^C=-e< out;{\bf p'}_N|
\biggl[A^C_{\mu}(0),{b}^+_{\bf {p_e}}(0)\biggr]\psi_e(0)|{\bf p}_N;in>=
-e{\overline u}({\bf p_e'})\gamma^o
\int { {d{\bf x'}}\over {4\pi|{\bf x'}|}}
<out;{\bf p}_N'|\psi_e^+(0,{\bf x'})\psi_e(0,0)
|{\bf p_N};in>u({\bf p_e})\eqno(3.10c)$$
The Poisson equation (3.5) and the equation of motion (3.4)
yield

$$Y_{I}^C={ {e{\overline u}({\bf p_e'})\gamma^ou({\bf p_e})}\over
 {-({\bf p_N'-p_N})^2} }<out;{\bf p}_N'|J^{tr}_o(0)|{\bf p}_N;in>
-e{\overline u}({\bf p_e'})\gamma^{i}u({\bf p_e}){1\over {t_N}}
<out;{\bf p}_N'|J^{tr}_{i}(0)|{\bf p}_N;in>
\eqno(3.11)$$

The longitudinal  part of 
$A_{i}^C$ and $J_{i}$ are excluded from $Y_{I}^C$ (3.11).
Nevertheless in the next section we shall demonstrate that $Y_{I}^C$ (3.11) 
coincides with  $V_{OPE}$ (2.7b).

$Y_{II}^C$   (3.10c) is nonlocal
due to integration of the nonlocal source
$e\psi^+_e(0,{\bf x'})\psi_e(0,{\bf 0})$.
Insertion of the completeness condition (2.4) 
between the electron fields in (3.10c) yields

$$Y_{II}^C=-e\sum_{n=e"N,e"N\gamma,...} 
{ {<out;{\bf p}_N'|{\overline \eta}_{\bf p_e'}(0)|n;in>_C }\over
{({\bf  p'_N-P_n})^2
\Bigl(i\gamma_{\sigma}(p_N'-P_n)^{\sigma} -m_e\Bigr)
 } }
{ {<in; n|\eta_{\bf p_e}(0)|{\bf p_N};in>}\over
{\Bigl(i\gamma_{\nu}(-p_N+P_n)^{\nu} -m_e\Bigr) }}
\eqno(3.12a)$$
$$+e\sum_{m=\gamma N,\gamma\gamma N,...} 
{ {<0|{\overline \eta}_{\bf p_e'}(0)|m;in>_C }\over
{{\bf P_m}^2
\Bigl(i\gamma_{\sigma}P_m^{\sigma} -m_e\Bigr)} }
{ {<in;{\bf p}_N', m|\eta_{\bf p_e}(0)|{\bf p_N};in>}\over
{\Bigl(i\gamma_{\nu}(p_N'-p_N-P_m)^{\nu} -m_e\Bigr) }}
\eqno(3.12b)$$
$$+e\sum_{m=\gamma N,\gamma\gamma N,..} 
{ {<out;{\bf p}_N'|{\overline \eta}_{\bf p_e'}(0)|m,{\bf p_N};in>_C }
\over{({\bf  p'_N-p_N-P_m})^2
\Bigl(i\gamma_{\sigma}(p_N'-p_N-P_m)^{\sigma} -m_e\Bigr)
 } }
{ {<in; m|\eta_{\bf p_e}(0)|0>}\over
{\Bigl(i\gamma_{\nu}P_m^{\nu}-m_e\Bigr) }}
\eqno(3.12c)$$
$$-e\sum_{ {\overline n}=e"{\overline N},e"{\overline N}\gamma,...} 
{ {<0|{\overline \eta}_{\bf p_e'}(0)|{\bf p_N}\ n;in>_C }\over
{({\bf  -p_N-P_{\overline n}})^2
\Bigl(i\gamma_{\sigma}(-p_N-P_{\overline n})^{\sigma} -m_e\Bigr) } }
{ {<in;  {\bf p}_N',{\overline n}|\eta_{\bf p_e}(0)|0>}\over
{\Bigl(i\gamma_{\nu}(p_N'+P_{\overline n})^{\nu} -m_e\Bigr) }}
\eqno(3.12d)$$

These  terms  have the same structure as the 
on mass shell particle
exchange amplitudes  (A.1a) - (A.1d) correspondingly.
The terms (3.12a)-(3.12d) consists of the products of the $ep$ scattering 
matrices and corresponding
crossing terms, i.e. these terms give $e^5$   contributions into 
$ep$ scattering amplitude. 
The nonlocality of (3.12a)-(3.12d) follows from the canonical equal time commutations  
between $b^+_{\bf p}(0)$ and the nonlocal field $A^C_o(0)$.

\vspace{0.25cm}

\begin{center}
{\bf{4. The Born term  (2.7a)  in
the Lorentz  and Coulomb gauges. }}
\end{center}

\vspace{0.25cm}

The Lorentz gauge 

$${{\partial {A^L}_{\mu}(x)}\over{\partial x_{\mu}}}=0\eqno(4.1)$$
allows to formulate
the  $ep$ scattering problem in the
Lorentz invariant 4D form if the four components of the photon
field ${{A^L}}_{\mu}$ are the independent variables.  
In order to avoid the problems with the indefinite metric 
in the Gupta-Bleuer formalism and simplify comparison with the
relations in the Coulomb gauge, 
we consider in the Lorentz gauge as independent fields
electrons and the transverse photons. 
Consequently we keep the canonical
equal time anti commutation (2.6c) for the electrons and
 (3.9) for the transverse part of   ${{A^L}}_{\mu}$. 
Then in the Lorentz gauge we get the following equation of motion

$$\Bigl(i\gamma^{\mu}{{\partial}\over{\partial x^{\mu}}}
-m_e\Bigr)\psi_e(x)=
\eta(x)= e\gamma^{\mu}
{A^L}_{\mu}(x)\psi_e(x)\eqno(4.2a)$$

$$\Box_x {(A^L)}^{tr}_{i}(x) 
=J^{tr}_i(x)=J_i(x)-
{{\partial}\over {\partial x^{i}}} 
 {{\partial J^{k}(x)}\over {\partial x^{k}}} ;\ \ \ i=1,2,3\eqno(4.2b)$$
  

and the zeroth and longitudinal components of the photon fields
are determined through the corresponding sources as

$$ {(A^L)}_{i}^l(x)
=\Box_x^{-1}J_{i}^l(x);\ \ \
J_{i}^l(x)=e{{\partial}\over {\partial x^{i}}} 
 {{\partial \Bigl[{\overline \psi_e}(x)\gamma_{k} \psi_e(x)\Bigr]
}\over {\partial x_{k}}}
\eqno(4.2c)$$

$$ {A^L}_{o}(x)
=\Box_x^{-1}J_{o}(x);\ \ \
J_{o}(x)=e{\overline \psi_e}(x)\gamma_{o} \psi_e(x)
,\eqno(4.2d)$$


where

$${(A^L)}_{i}^{tr}(x)={A^L}_{i}(x)-{{\partial}\over {\partial x_{i}}}
 {{\partial {A^L}_{k}(x)}\over {\partial x_{k}}};\ \ \ \ \
{(A^L)}_{i}^{l}(x)={{\partial}\over {\partial x^{i}}}
 {{\partial {(A^L)}_{k}(x)}\over {\partial x_{k}}}
\eqno(4.3)$$

The gauge condition (4.1) allows to redefine  ${(A^L)}_{i}^{l}$
 through the ${A^L}_{o}$  

$${(A^L)}_{i}^{l}(x)={{\partial}\over {\partial x^{i}}}
 {{\partial {A^L}_o(x)}\over {\partial x^{o}} }
\eqno(4.4)$$

Thus we have two auxiliary fields
$({A^L})^{l}_{i}$ and ${A^L}_{o}$ 
in the Lorentz gauge (4.1). Both of these  auxiliary fields
are determined through the $J_{o}$ according to 
 (4.2c,d) and (4.4).

The first part of the equal time commutator (2.7a) 
$Y^l=Y_{I}^L-Y_{II}^L$  in the Lorentz gauge

$$Y^{L}_{I}=
e{\overline u}({\bf p_e'})\gamma^{\mu    }
< out;{\bf p'}_N|{A^L}_{\mu}(0)
\biggl\{\psi_e(0),b_{\bf p_e}^+(0)\biggr\}|{\bf p}_N;in>=
e{\overline u}({\bf p_e'})\gamma^{\mu}u({\bf p_e})
< out;{\bf p'}_N|{A^L}_{\mu}(0)|{\bf p}_N;in>\eqno(4.5)$$
 reproduces exactly the  Born term $V_{OPE}$ (2.7b).

The next part of the equal-time commutators (2.7a)

$$Y^{L}_{II}=
-e{\overline u}({\bf p_e'})\gamma^{\mu=0,3}
< out;{\bf p'}_N|\biggl[{A^L}_{\mu=0,3}(0),{b}^+_{\bf {p_e}}(0)\biggr]\psi_e(0)
|{\bf p}_N;in>\eqno(4.6)$$
is more complicated  than  $Y^C_{II}$ (3.10c)  
 because ${A^L}_{\mu=0,3}$ in (4.2d)
contains additional integration over the $x_o'$ 
of the sources $J_o(x')$. 

Thus $Y^{L}_{I}$ (4.5) and $Y^{C}_{I}$ (3.11) determine the leading one photon
 exchange (OPE) term  in the Lorentz and Coulomb gauge correspondingly. These 
OPE potentials (Fig. 1) are constructed via the nucleon vertexes
$$F^L_{\mu}=<out; {\bf p}_N'|{A^L}_{\mu}(0)|{\bf p}_N;in>\eqno(4.7)$$
$$F^C_{\mu}=<out; {\bf p}_N'|{A^C}_{\mu}(0)|{\bf p}_N;in>.\eqno(4.8)$$
The photon fields  in the Lorentz and Coulomb gauges
${A^L}_{\mu}(x)$ and  ${A^C}_{\mu}(x)$ 
can be obtained independently
through the gauge  transformation in the 
Dirac equation for the noninteracting electrons.
In order to determine the relationship between the vertexes 
(4.7) and (4.8) 
we  consider  the gauge transformations 
between  these fields
$$\Psi^C(x)=e^{i\lambda(x)}\Psi^L(x);\ \ \ \ \ \ \ \ \ \ \ \ \ \ \ \ \ \ \ \ \ \ 
{ A}^C_{\mu}(x)=e^{-i\lambda(x)}{ {A^L}}_{\mu}(x)e^{i\lambda(x)}
+{{\partial {\lambda}(x)}\over {\partial x_{\mu}}}
\eqno(4.9)$$
where the electron field  $\psi_e$ is denoted as $\Psi^L$ and
$\Psi^C$ in the Lorentz and  Coulomb gauges.

Generally $\lambda(x)$  contain the auxiliary 
fields ${A^L}_{\mu=3,0}$ which does not commute with
 $\Psi^L$. Therefore, it is
 convenient to extract from ${  A}^C_{\mu}$ the  part 

$${\cal D}_{\mu}(x)\equiv
e^{-i\lambda(x)}{{A^L}}_{\mu}(x)e^{i\lambda(x)}-{A^L}_{\mu}(x)=
\Bigl[{A^L}_{\mu=0,3}(x),\lambda(x)\Bigr]+
\biggl[\Bigl[{A^L}_{\mu=0,3}(x),\lambda(x)\Bigr],\lambda(x)\biggr]+...\eqno(4.10)$$

and chose $\lambda$ via the solution of the equation

$${{\partial {\lambda}(x)}\over {\partial x_{\mu}}}+{\cal D}_{\mu}(x)=
-{1\over {\Delta}}
{{\partial }\over {\partial x_{\mu} } }
{{\partial {A^L}_{o}(x)}\over {\partial x_o} }\eqno(4.11)$$

Then we get

$$A_{\mu}^C(x)={A^L}_{\mu}(x)-{1\over {\Delta}}
{{\partial }\over {\partial x_{\mu} } }
{{\partial {A^L}_{o}(x)}\over {\partial x_o} }
\eqno(4.12)$$

which is valid also in the classical physic,
where $\lambda$ commute with $A_{\mu}$ and ${\cal D}_{\mu}=0$.

It is easy to check that $A^C_{\mu}$ (4.12) satisfy
the gauge conditions (3.1),
the equation of motion (3.3) and the Poisson equation (3.5). 
Moreover,   $\Psi^C$ after gauge transformation (4.9) satisfies 
the equation of motion (3.4).

 Substitution $A_{\mu}^C$ (4.12) in  (3.5) yields
$$-{\Delta}A^C_o=\Box {A^L}_o=J_o(x)\eqno(4.13)$$
which produces the following relationship between the nucleon
form factors in the Lorentz and Coulomb gauges

$$t_N<out;{\bf p'}_N|A^L_{o}(0)|{\bf p}_N;in>=-({\bf p'_N-p_N})^2
<out;{\bf p'}_N|A^C_{o}(0)|{\bf p}_N;in>\eqno(4.14)$$

The relation (4.12) allows equate 
 the leading one photon exchange
potential $V_{OPE}$  (2.7b) in the Lorentz and Coulomb gauges 
$$V_{OPE}^L=V_{OPE}^C\eqno(4.15a)$$ 
where
$$V_{OPE}^L
=e{\overline u}({\bf p_e'})\gamma_{\mu}u({\bf p_e})
<out;{\bf p'}_N|A^L_{\mu}(0)|{\bf p}_N;in>WY,\eqno(4.14b)$$
$$V_{OPE}^C
=e{\overline u}({\bf p_e'})\gamma_{\mu}u({\bf p_e})
<out;{\bf p}_N'|A^C_{\mu}(0)|{\bf p}_N;in>,\eqno(4.14c)$$
because
$${\overline u}({\bf p_e'})\gamma_{\mu}u({\bf p_e})
<out;{\bf p}_N'|\biggl[
{{\partial }\over {\partial x_{\mu} } }
{{\partial {A^L}_{o}(x)}\over {\partial x_o} }\biggr]_{x=0}
|{\bf p}_N;in>=$$
$$-i{\overline u}({\bf p_e'})\gamma_{\mu}(p_N'-p_N)^{\mu}u({\bf p_e})
<out;{\bf p}_N'|\biggl[
{{\partial {A^L}_{o}(x)}\over {\partial x_o} }\biggr]_{x=0}
|{\bf p}_N;in>
=0\eqno(4.16)$$

The canonical equal time anti commutator
of the electron fields in the Lorentz gauge

$$\Bigl\{ {\Psi^{L}}^+(x_o,{\bf x}),\Psi^{L}(x_o,{\bf y})\Bigr\}=
\delta({\bf x-y})\eqno(4.17a)$$ 

after gauge transformation (4.9)  
produces the following canonical anti commutators in the Coulomb gauge

$$\Bigl\{{\Psi^{C}}^+(x_o,{\bf x}),\Psi^C(x_o,{\bf y})\Bigr\}=
\delta({\bf x-y})+ {\cal C}(x,{\bf y}),\eqno(4.17b)$$

where

$${\cal C}(x,{\bf y})=
{\Psi^L}^+(x)\Bigl[e^{-i\lambda(x)},e^{i\lambda(x_o.{\bf y})}\Bigl]
\Psi^L(x_o,{\bf y})$$
$$+\Bigl[{\Psi^L}^+(x),e^{i\lambda(x_o.{\bf y})}\Bigl]e^{-i\lambda(x_o.{\bf y})}{\Psi^L}(x_o,{\bf y})
+e^{i\lambda(x_o.{\bf y})}\Bigl[{\Psi^L}(x_o,{\bf y}),e^{-i\lambda(x_o.{\bf y})}\Bigl]{\Psi^L}^+(x)
\eqno(4.18)$$

The anti commutator  (4.17) and the gauge transformation (4.9) 
generate the additional terms
with ${\cal D}_{\mu}$ (4.10)  and ${\cal C}_{\mu}$ (4.18)
in $Y^C_I$ (3.10b) and  $Y^C_{II}$ (3.10c) which gives also the small 
corrections in order to $e^5$.

\vspace{0.25cm}
\begin{center}
{\bf{5. Conclusion}}
\end{center}
\vspace{0.25cm}

In this paper  the general time ordered field-theoretical 
unitarity condition (2.5) are used for derivation
of the 3D relativistic
Lippmann-Schwinger type equations (2.13) for the $ep$ scattering amplitude
${\overline u}({\bf p_e}')<out;{\bf p'}_N|\eta(0)|
{\bf p}_e{\bf p}_N;in>$ (2.1a). In the unitarity condition (2.5)
and the corresponding equation (2.13) the nucleons are on mass shell 
and  only one of the external electrons is off mass shell.
Therefore in this approach  the proton self energy diagrams, 
the proton vertex correction and other diagrams  with the off mass shell 
nucleons do not appear. The leading order part of the potential  of 
the equation (2.13) is the  one photon 
exchange  potential $V_{OPE}$ (2.7b) in Fig. 1 which is constructed via 
the usual one variable electromagnetic form factors of the nucleon (2.9).

The unitarity condition (2.5)  follows directly from the 
${\cal S}$-matrix reduction 
formula for the $ep$ scattering amplitude ${\cal A}_{e'N',eN}$ (2.1a). Consequently.
 unitarity condition (2.5) and it representation (2.13) 
are the necessary conditions for the $ep$ scattering amplitude in $QED$.  
Moreover, the ${\cal S}$-matrix reduction technique provides us
with  the analytic relations between the 
${\cal A}_{e'N',eN}$ (2.1a) and the corresponding 4D
amplitude of the Bethe-Salpeter equation and 
their 3D quasipontial reductions.
Therefore any results  obtained in the present approach can be reproduced
within the 4D formulation and vice versa.




It is shown that the leading order potential $V_{OPE}$ (2.7b) together with the additional
terms (3.10c) are generated by the canonical equal time anti commutators (2.7a).
These anti commutators are not invariant under the gauge 
transformations of the commuting fields.  
Nevertheless
it is demonstrated that $V_{OPE}$ is same in the Coulomb and 
Lorentz gauge. 
The additional terms (3.10c) and (3.12a)-(3.12d) are 
 produced by the electrostatic (Coulomb) interactions and they
give small corrections in order of $\sim e^5$. 
The Coulomb gauge is more convenient for calculation of these corrections.

The   unitarity condition (2.5)  contains only the hadron and lepton degrees of 
freedom due to the equal time commutators of the electron fields 
(2.7a) and the completeness
condition of the asymptotic particle states (2.4). 
Therefore the quark-gluon degrees 
of freedom can contribute in (2.5) and (2.13) only through
the input electromagnetic nucleon vertex 
$<out; {\bf p'}|J_{\mu}(0)|{\bf p};in''$ (2.9).

The alternative unitarity condition can be obtained 
using  the ${\cal S}$-matrix reduction formulas for the on mass shell 
electrons and off mass shell nucleons. The composed nucleon field  
can be constructed within the Haag-Nischijima-Zimmermann approach
\cite{H,N,Z,HW}.
In this case the equal time commutators for the composed nucleon fields
\cite{M,MBF,MFJPG}
  as the quark bound state operators 
 contains all contributions from the quark-gluon degrees of freedom
and  the corresponding  Lippmann-Schwinger type equation  generates the
electromagnetic corrections of this vertex.


\vspace{0.25cm}

\begin{center}
{\bf{Appendix A. Linearization of the generalized unitarity condition (2.6).}}
\end{center}

\vspace{0.25cm}

Separation of the connected and disconnected parts
of the amplitudes  in  two last
terms of (2.5) yields

$${\sc w}_{e'N',eN}(n)=(2\pi)^3\sum_{n}
<out;{\bf p'}_N|\eta_{\bf p_e'}(0)|n;in>_C
{{\delta({\bf P_n-p_e-p_N})}\over {P_o-P_n^o+i\epsilon}}
<in;n|{\overline \eta}_{\bf p_e}(0)|{\bf p}_N;in>_C\eqno(A.1a)$$

$$
-(2\pi)^3 \sum_{m=\gamma N,\gamma\gamma N,...} < 0 |\eta_{\bf p_e'}(0)|m;in>{\frac{ {%
\delta^{(3)}( {\bf p}_e+{\bf p}_{2}-{\bf P}_{m}-{\bf p^{\prime}}_{2} )}}{{%
P_o-P_{m}^o-E_{{\bf p^{\prime}}_{2} }
+i\epsilon } }} <in;{\bf p^{\prime}}_{2},m|{\overline \eta}_{\bf p_e}(0)|{\bf p}_{2};in>_c%
\eqno(A.1b)
$$

$$
-(2\pi)^3 \sum_{m=\gamma N,\gamma\gamma N,...} <in;{\bf p^{\prime}}_N |\eta_{\bf p_e'}(0)|%
{\bf p}_{2},m;in>_c {\frac{ {\delta^{(3)}( {\bf p}_e-{\bf P}_{m} )}}{{%
E_{{\bf p_e}}-P_{m}^o } }} <in;m|{\overline \eta}_{\bf p_e}(0)|0>\eqno(A.1c)
$$

$$
-(2\pi)^3 \sum_{{\overline n}=e''{\overline N},\gamma e''{\overline N}...} 
<0 |\eta_{\bf p_e'}(0)|{\bf p}_{2}{\overline n};in> 
{ {\delta^{(3)}( {\bf p}_e-{\bf p^{\prime}}_{2}-{\bf P}_{\overline n} )}\over
{E_{\bf p_e}- E_{{\bf p^{\prime}}_{2}}-P^o_{\overline n } }}
<in;{\bf p_N'} {\overline n}|{\overline \eta}_{\bf p_e}(0)|0>\eqno(A.1d)
$$


$$-(2\pi)^3\sum_{\upsilon={\overline e}N,\gamma{\overline e} N...}
<out;{\bf p'}_N|{\overline \eta}_{\bf p_e}(0)|\upsilon;in>_C
{{\delta({\bf P_{\upsilon}+p_e-p_N'})}\over {
-E_{\bf p_N'}+E_{\bf p_e}+P_{\upsilon}^o}}
<in;\upsilon| \eta_{\bf p_e'}(0)|{\bf p}_N;in>_C,   
\eqno(A.1e)$$

$$+(2\pi)^3\sum_{{\overline m}={\overline e},\gamma{\overline e}...}
<0|{\overline \eta}_{\bf p_e}(0)|{\overline m};in>
{{\delta({\bf P_{\overline m}+p_e})}\over {
E_{\bf p_e}+P_{\overline m}^o}}
<in;{\bf p'}_N{\overline m}| \eta_{\bf p_e'}(0)|{\bf p}_N;in>_C,   
\eqno(A.1f)$$

$$+(2\pi)^3\sum_{{\overline m}={\overline e},\gamma{\overline e}...}
<out;{\bf p'}_N|{\overline \eta}_{\bf p_e}(0)
|{\overline m}{\bf p}_N;in>_C
{{\delta({\bf P_{\overline m}+p_e+p_N-p_N'})}\over {
-E_{\bf p_N'}+E_{\bf p_e}+E_{\bf p_N}+P_{\overline m}^o}}
<in;{\overline m}| \eta_{\bf p_e'}(0)|0>,   
\eqno(A.1g)$$

$$+(2\pi)^3\sum_{{\overline n}={\overline e}{\overline N},\gamma{\overline e}{\overline N}...}
<0|{\overline \eta}_{\bf p_e}(0)|{\overline n}{\bf p}_N;in>
{{\delta({\bf P_{\overline n}+p_e+p_N})}\over {
E_{\bf p_e}+E_{\bf p_N}+P_{\overline n}^o}}
<in;{\overline n}{\bf p'}_N| \eta_{\bf p_e'}(0)|0>,   
\eqno(A.1h)$$
where $n$ denotes
the complete set of the
$s$-channel intermediate states $n=H,eN,\gamma eN,\gamma\gamma eN,\gamma H...$.
The index $_C$ indicates the connected part of the related amplitude and the factor $\pm$ of the expressions (A.1b)-(A.1h) appears after transposition of the fermion fields of the electron and nucleons. 

${\sc w}_{e'N',eN}(n)$   without $s$-channel $H$ and $eN$ exchange terms forms the
inhomogeneous part ${\cal  W}_{e'N',eN}$ of the relation (2.10)   
$${\sc w}_{e'N',eN}(n=\gamma eN,\gamma\gamma eN,\gamma H...)={\cal  W}_{e'N',eN}\eqno(A.2)$$

${\cal  W}_{e'N',eN}$ with the minimal number of the intermediate states
are depicted in Fig. 1A - Fig. 1H correspondingly.
It is easy to see, that The last four terms in Fig. 1E-Fig. 1H are
obtained from the  terms in Fig. 1A-Fig. 1D by crossing of the 
electrons.

After  simple algebra for the  propagators in (A.1a)-(A.1h) we get

$${\cal W}_{e'N',eN}({\bf p',p})=
{\sf A}_{e'N',eN}({\bf p',p})+
P_o'{\sf B}_{e'N',eN}({\bf p',p}),\eqno(A.3)$$
where ${\sf A}_{e'N',eN}$ and $={\sf B}_{e'N',eN}$ 
are the following Hermitian matrices 
$${\sf A}_{e'N',eN}
=(2\pi)^3\sum_{n=\gamma eN,\gamma\gamma eN,\gamma H...}
<out;{\bf p'}_N|\eta_{\bf p_e'}(0)|n;in>_C
{{(-P_n^o)\delta({\bf P_n-p_e-p_N})}\over {P_o-P_n^o+i\epsilon}}
{{<in;n|{\overline \eta}_{\bf p_e}(0)|{\bf p}_N;in>_C}
\over {P_o'-P_n^o-i\epsilon}}$$

$$
-(2\pi)^3 \sum_{m=\gamma N,\gamma\gamma N,...} < 0 |\eta_{\bf p_e'}(0)|m;in>{
\frac{(-P_{m}^o-E_{{\bf p'}_{2} }) {%
\delta^{(3)}( {\bf p}_e+{\bf p}_{2}-{\bf P}_{m}-{\bf p^{\prime}}_{2} )}}{{%
P_o-P_{m}^o-E_{{\bf p^{\prime}}_{2} }
+i\epsilon } }} 
{{<in;{\bf p^{\prime}}_{2},m|{\overline \eta}_{\bf p_e}(0)|{\bf p}_{2};in>_c}\over
{P_o'-P_{m}^o-E_{{\bf p'}_{2} }}}
$$

$$
-(2\pi)^3 \sum_{m=\gamma N,\gamma\gamma N,...} 
<in;{\bf p^{\prime}}_N |\eta_{\bf p_e'}(0)|{\bf p}_{2},m;in>_c 
{\frac{(-P_{m}^o-E_{{\bf p}_{2} }) {\delta^{(3)}( {\bf p}_e-{\bf P}_{m} )}}
{{E_{{\bf p_e}}-P_{m}^o } }} {{<in;m|{\overline \eta}_{\bf p_e}(0)|0>}
\over {P_o'-P_{m}^o-E_{{\bf p}_{2} } -i\epsilon}}
$$

$$
=(2\pi)^3 \sum_{{\overline n}=e''{\overline N},\gamma e''{\overline N}...} 
<0 |\eta_{\bf p_e'}(0)|{\bf p}_{2}{\overline n};in> 
{ {(-E_{\bf p_N}- E_{{\bf p^{\prime}}_{2}}-P^o_{\overline n })
\delta^{(3)}( {\bf p}_e-{\bf p^{\prime}}_{2}-{\bf P}_{\overline n} )}\over
{E_{\bf p_e}- E_{{\bf p^{\prime}}_{2}}-P^o_{\overline n } }}
{{<in;{\bf p_N'} {\overline n}|{\overline \eta}_{\bf p_e}(0)|0>}
\over {P_o'-E_{\bf p_N}- E_{{\bf p^{\prime}}_{2}}-P^o_{\overline n }}}$$
$$+electron\ \ \  crossing\ \ \  terms\eqno(A.4)$$

$${\sf B}_{e'N',eN}=(2\pi)^3\sum_{n=\gamma eN,\gamma\gamma eN,\gamma H...}
{{<out;{\bf p'}_N|\eta_{\bf p_e'}(0)|n;in>_C}\over
{P_o'-P_n^o-i\epsilon  }}
\delta({\bf P_n-p_e-p_N})
{{<in;n|{\overline \eta}_{\bf p_e}(0)|{\bf p}_N;in>_C}
\over {P_o-P_n^o+i\epsilon}}$$

$$
-(2\pi)^3 \sum_{m=\gamma N,\gamma\gamma N,...} 
{{< 0 |\eta_{\bf p_e'}(0)|m;in>}\over{P_o'-P_{m}^o-E_{{\bf p'}_{2} } }}
\delta^{(3)}( {\bf p}_e+{\bf p}_{2}-{\bf P}_{m}-{\bf p^{\prime}}_{2} )
  {{<in;{\bf p^{\prime}}_{2},m|{\overline \eta}_{\bf p_e}(0)|{\bf p}_{2};in>_c}
\over{P_o-P_{m}^o-E_{{\bf p^{\prime}}_{2} }+i\epsilon }}
$$

$$
-(2\pi)^3 \sum_{m=\gamma N,\gamma\gamma N,...} 
{{<in;{\bf p^{\prime}}_N |\eta_{\bf p_e'}(0)|{\bf p}_{2},m;in>_c}\over
{P_o'-P_{m}^o-E_{{\bf p}_{2} } -i\epsilon}} \delta^{(3)}( {\bf p}_e-{\bf P}_{m} )
  {{<in;m|{\overline \eta}_{\bf p_e}(0)|0>}\over {E_{{\bf p_e}}-P_{m}^o}}
$$

$$
=(2\pi)^3 \sum_{{\overline n}=e''{\overline N},\gamma e''{\overline N}...} 
{{<0 |\eta_{\bf p_e'}(0)|{\bf p}_{2}{\overline n};in>}
\over {E_{{\bf p^{\prime}}_e}-E_{\bf p_N}-P^o_{\overline n } }}
\delta^{(3)}( {\bf p}_e-{\bf p^{\prime}}_{2}-{\bf P}_{\overline n} )
{{<in;{\bf p_N'} {\overline n}|{\overline \eta}_{\bf p_e}(0)|0>}
\over {E_{\bf p_e}- E_{{\bf p^{\prime}}_{2}}-P^o_{\overline n }} }
$$
$$+\ electron\ \ \ crossing\ \ \ 4\ \ \ terms\eqno(A.5)$$

The Hermitian matrices ${\sf A}_{e'N',eN}$ (A.4) and 
${\sf B}_{e'N',eN}$ (A.5) together with the 
equal-time commutator $Y_{e'N',eN}$ (2.7a) form the 
linear energy depending potential 
${\cal U}({\bf p',q},E)$ 
of the relativistic Lippmann-Schwinger equation (2.13)

$${\cal U}({\bf p',q},E)\equiv {\cal U}(E)=
Y_{e'N',eN}+{\sf A}_{e'N',eN}+E{\sf B}_{e'N',eN}\eqno(A.6)$$

It must be noted that  one can determinate ${\sf B}_{e'N',eN}$ (A.5) 
through the $s$ channel terms with the $H$ and $eN$ intermediate states

$$\delta^{(3)}({\bf P'-P})\biggl[
(2\pi)^3\sum_{n=H;e"N"}
<out;{\bf p'}_N|\eta_{\bf p_e'}(0)|n;in>_C
{{\delta({\bf P_n-p_e-p_N})}\over {P_o-P_n^o+i\epsilon}}
<in;n|{\overline \eta}_{\bf p_e}(0)|{\bf p}_N;in>_C$$
$$+{\sf B}_{e'N',eN}\biggr]
=<in; {\bf p_N'}|\Bigl\{ b_{\bf p_e}(0), b^+_{\bf p_e}(0)\Bigr\}
|{\bf p_N };in>
=<in;{\bf p_e'p_N'}|{\bf p_e p_N};in>\eqno(A.7)$$ 
Therefore, the operator $(1-B)$ corresponds to 
the contributions of the
intermediate $s$ channel $eN$ and $H$ states in the 
commutation relation for the Heisenberg field operators.

Using the same technique as 
for the other binary reactions in \cite{MR,MCH,M,MBF,MF}.
one can derive  (2.5) from the Lippmann-Schwinger-type 
equation (2.13) and vice verse


\end{document}